\documentclass[12pt]{article}
\usepackage{verbatim,color,amssymb,bbm,epsfig}
\usepackage[usenames,dvipsnames]{xcolor}
\usepackage{fancyhdr}
\usepackage[authoryear, sort]{natbib}
\usepackage{wasysym}
\usepackage{float}
\def\K{{\cal K}}

\def\boxit#1{\vbox{\hrule\hbox{\vrule\kern6pt  \vbox{\kern6pt#1\kern6pt}\kern6pt\vrule}\hrule}}

\def\sumi{\hbox{$\sum_{i=1}^n$}}

\def\wh{\widehat}

\def\bse{\begin{eqnarray*}}
	\def\ese{\end{eqnarray*}}
\def\be{\begin{eqnarray}}
\def\ee{\end{eqnarray}}
\def\bq{\begin{equation}}
\def\eq{\end{equation}}
\def\bse{\begin{eqnarray*}}
	\def\ese{\end{eqnarray*}}
\def\pr{\hbox{pr}}
\def\wh{\widehat}
\def\trans{^{\rm T}}

\def\th{^{th}}
\def\b1e{{\mathbf e}}

\def\bX{{\mathbf X}}

\newcommand{\bbeta}{\mbox{\boldmath $\beta$}}

\def\boldbeta{\boldmath\beta}

\def\boldtheta{\boldmath\theta}

\def\trans{^{\rm T}}

\def\th{^{th}}
\def\b1e{{\mathbf e}}

\def\bX{{\mathbf X}}

\def\bZ{{\mathbf Z}}

\def\sumb{\hbox{$\sum_{b=1}^{B}$}}

\def\doutb{${\cal D}_{out}^{(b)}$}
\def\dinb{${\cal D}_{in}^{(b)}$}
\def\Supp{Supplementary Material}

\renewcommand\footnoterule{\kern-3pt \hrule \textwidth 2in \kern 2.6pt}

\def\boxit#1{\vbox{\hrule\hbox{\vrule\kern6pt \vbox{\kern6pt \textcolor{blue}{#1}\kern6pt}\kern6pt\vrule}\hrule}}

\def\authorfootnote#1{{\let\thefootnote\relax\footnotetext{#1}}}

\begin{document}
	
	\thispagestyle{empty}
	\baselineskip=28pt
	\begin{center}
	{\LARGE{\bf Sample Splitting as an M-Estimator with Application to Physical Activity Scoring}}
	\end{center}
	\baselineskip=12pt
	
	\vskip 2mm
	\begin{center}
		Eli. S. Kravitz\\
		Department of Statistics, Texas A\&M University, 3143 TAMU, College Station, TX 77843-3143, USA, kravitze@tamu.edu\\
		\hskip 5mm \\
		Raymond J. Carroll\\
		Department of Statistics, Texas A\&M University, 3143 TAMU, College Station, TX 77843-3143, USA and School of Mathematical and Physical Sciences, University of Technology Sydney, Broadway NSW 2007, Australia, carroll@stat.tamu.edu\\
		\hskip 5mm \\
		David Ruppert\\
		School of Operations Research and Information Engineering and Department of Statistics and Data Science, Cornell University, Ithaca NY 14853, USA, dr24@cornell.edu
	\end{center}

	\vskip 10mm
	\begin{center}
		{\Large{\bf Abstract}}
	\end{center}
	
	\baselineskip=16pt
	Sample splitting is widely used in statistical applications, including classically in classification and more recently for inference post model selection. Motivating by problems in the study of diet, physical activity, and health, we consider a new application of sample splitting. Physical activity researchers wanted to create a scoring system to quickly assess physical activity levels. A score is created using a large cohort study. Then, using the same data, this score serves as a covariate in a model for the risk of disease or mortality. Since the data are used twice in this way, standard errors and confidence intervals from fitting the second model are not valid.
	To allow for proper inference, sample splitting can be used. One builds the score with a random half of the data and then uses the score when fitting a model to the other half of the data.  We derive the limiting distribution of the estimators. An obvious question is what happens if multiple sample splits are performed. We show that as the number of sample splits increases, the combination of multiple sample splits is effectively equivalent to solving a set of estimating equations.
	
\baselineskip=12pt
\par\vfill\noindent

\par\medskip\noindent
\underline{\bf Short title}: Sample Splitting as an M-Estimator

\clearpage\pagebreak\newpage
\pagenumbering{arabic}
\newlength{\gnat}
\setlength{\gnat}{22pt}
\baselineskip=\gnat

\section{Introduction} \label{sec:introduction}
	
In many applications it is useful to think of performing a statistical procedure in two stages: model building and validation. When building a classification model, it is commonplace to reserve a portion of the data set for testing predictive accuracy. Similarly, many nonparametric regression methods depend on an unknown regularization parameter, $\lambda$, which may be chosen using cross-validation.  Here, model fit is determined using data which was not used to build the model. In both examples, reusing data for model fitting and validation without withholding a portion will result in over-fitting.

This two stage procedure can be seen in real-world problems. Consider providing dietary recommendations in the form of \textit{composite score} or \textit{index}. One particular example is the Healthy Eating Index. The Healthy Eating Index was designed by the United States Department of Agriculture (USDA) to provide dietary recommendations to Americans \citep{hei_usda}. The Healthy Eating Index was later validated and confirmed to accurately measure diet and predict risk of disease with a 24-hour food recall study \citep{guenther2008evaluation}.

In Section \ref{sec:motivation}, we introduce a scenario of building a composite score and providing risk estimates without a second independent population. We split the data set, $\mathcal{D}$, into two pieces, $\mathcal{D}_{in}$ and $\mathcal{D}_{out}$. The partition, $\mathcal{D}_{in}$, serves as a ``training" data set for model fitting while $\mathcal{D}_{out}$ serves as a ``validation" data set for checking model fit and performing hypothesis tests. In this way we can mimic  building and validation with two independent populations.

It is natural to extend this process: If one split is good, \textit{many splits} should be better. A single split has variability; a fortuitous split may give parameter estimates very close to their true value while another split may yield poor estimates. We follow the ideas of \cite{meinshausen2009}, though in a different context than in their variable selection paper. These authors create $B$ copies of the data and partition the $b$th copy into $\mathcal{D}^{(b)}_{in}$ and $\mathcal{D}^{(b)}_{out}$. For each $b$, model fitting is done on $\mathcal{D}^{(b)}_{in}$ and hypothesis tests are performed on $\mathcal{D}^{(b)}_{out}$. 

We explore multiple splitting for parametric models in Section \ref{sec:methodology}. We fit a model with parameter $\theta$ to \dinb\ and then fit a model with parameter $\beta_0$ to \doutb\  using a function of $\widehat \theta$ as a predictor. Using sample splitting, \cite{kravitz2019a} developed an exact test of $H_0: \beta_0 = 0$.
Exact confidence intervals for $\beta_0$ are not available, but, instead, we develop a large-sample confidence interval.
We find the surprising result that as the number of splits and the sample size approach $\infty$, the split-sample estimators become equivalent to fitting both models simultaneously using \textit{stacked estimating equations} \citep[Appendix A.6.6]{carroll2006measurement}.

Our method is similar to \textit{divide-and-conquer algorithms} \citep{li2013statistical, battey2015distributed}, particularly the average mixture algorithm  \citep{chang2017divide, zhang2012communication}. These algorithms reduce the computational cost of statistical inference on large datasets by dividing the dataset into many smaller subsets which can be quickly analyzed with traditional software (R, SAS, etc). The results from these subsets are then combined together, often with the sample mean.  None of this work, to our knowledge, incorporates a dependent parameter and fits a second model to estimate this parameter. 

This paper is organized as follows: In Section~\ref{sec:motivation}, we introduce a motivating example. In Section~\ref{sec:methodology} we describe sample splitting using estimating equations and M-estimators.  Section~\ref{sec:asymptotics} shows theoretical results for a single split, a finite number of splits, and when the number of splits approach infinity.  Section~\ref{sec:simulations} has simulations that investigate finite-sample behavior.  Section~\ref{sec:data_analysis} uses sample splitting to develop a physical activity index. Section~\ref{discuss} has a discussion. All technical details are collected in an Appendix.

\section{Motivating Example: Physical Activity and Survival} \label{sec:motivation}

This work is partly motivated by the creation and analysis of a physical behavior score to predict mortality. Researchers may be concerned about using data to create a score and then using the same date to validate the score as a predictor of risk.  We have found sample splitting appealing to practitioners because the relative risks are estimated with a different dataset than that which builds the score.

We build the physical behavior score using the NIH-AARP Study of Diet \citep{schatzkin2001design}. Participants self-reported physical behaviors which are then characterized into 8 discrete components. 
A priori, we specify the expected relationships. e.g., concave increasing, between these aerobic activity and survival to be consistent with the kinesiology literature.
Using the training data $\mathcal{D}_{in}$ for a single sample split,  we fit a binary regression model to survival that satisfies these relationships. The 8 components and their marginal models are listed in Table \ref{constraints}. The expected relationships are justified in Section \ref{sec:data_analysis}. We  rescale the fitted from this model so people with high levels of beneficial activity are assigned a score near 100 and people with low levels of beneficial activity are assigned a score close to 0. We denote the rescaled fitted values as $f(\bX; \wh{\theta})$

We now ask: Is our composite  score predictive of mortality? If so, how strong is the effect? We use $f(\bX; \wh{\theta})$ as a predictor in a logistic regression, along with covariates, $\bZ$. In the test data $\mathcal{D}_{out}$, we the model the probability of mortality as
\be \label{intro_cox}
\pr(Y_i = 1 \vert \bX_i,Z_i) =  H\{ \beta_0  f(\bX_i; \wh{\theta}) + \bZ_i \trans \beta \},
\ee
where $H(\cdot)$ is the logistic distribution function. Our primary interest is in $\beta_0$, which describes the relative risk of survival as a function of the composite score.

Two-stage modeling is convenient and therefore appealing to practitioners. 
In our application, model $f(\bX,\wh{\theta})$ was complex and required that imposition of shape constraint to be consistent with the kinesiology literature, so it was essential that we could fit that model first without needing to fit (\ref{intro_cox}) simultaneously.
Also, with $\wh \theta$  fixed, (\ref{intro_cox}) is a logistic model and can be fit with standard software.

\section{Sample Splitting Methodology} \label{sec:methodology}
Denote the total number of sample splits as $B$. We consider two cases: a single sample split $(B = 1)$ and many splits $(1 < B < \infty)$. We describe the algorithm for sample splitting using estimating equations. In Section \ref{sec:asymptotics}, we provide an asymptotic theory for both of these cases, as well when $B \to \infty$.

\subsection{Basic Formulation} \label{sec:basics}

In what follows, we use the term {\it estimating equation} to also include a {\it M-estimator equation}. We observe two types of responses $(W_i,Y_i)$ that are independent and identically distributed and two types of covariates, $(\bX_i,\bZ_i)$. There are two parameters,  $\boldtheta$ and $\boldbeta$. The relationship between $W$ and $(\bX,\bZ,\boldtheta)$ is described by a model which has an estimating function $\Psi(W,\bX,\bZ,\boldtheta)$.  The relationship between $Y$ and $(\bX,\bZ,\boldbeta,\boldtheta)$ is described by a model that has an estimating function $\K(Y,\bX,\bZ,\boldbeta,\boldtheta)$. We define $\boldtheta$ and $\boldbeta$ as the solutions to $E\{ \Psi(W,\bX,\bZ,\boldtheta)\}=0$ and $E\{\K(Y,\bX,\bZ,\boldbeta,\boldtheta)\}=0$ respectively.

For a single data set, consistent estimation of $(\boldbeta,\boldtheta)$ can be done routinely by solving the stacked estimating equation
\be
0 &=& \sumi \Big\{\Psi\trans(W_i,\bX_i,\bZ_i,\boldtheta),\K\trans(Y_i,\bX_i,\bZ_i,\boldbeta,\boldtheta)\Big\}. \label{eq01}
\ee
It is assumed that $\sumi \Psi(W_i,\bX_i,\bZ_i,\boldtheta)=0$ has  unique solution $\theta$, so in the second equation, $\sumi \K (Y_i,\bX_i,\bZ_i,\boldbeta,\boldtheta)=0 $, $\theta$ is fixed and the solution is in $\beta$.
Asymptotic theory for such estimators is well-known \citep{huber1964robust, huber1967behavior,stefanski2002calculus}.

In the physical activity problem of Section~\ref{sec:motivation}, $\K(\cdot)$ does not depend directly on $\theta$ but rather on a function of $\theta$ and $\bX$, denoted with $f(\bX; \theta)$. That is,
\bse
\K(Y,\bX,\bZ,\boldbeta,\boldtheta) &=& \K(Y,f(\bX; \boldtheta),\bZ,\boldbeta). \label{eq02}
\ese
Stacked estimating equations and the methods in this paper apply to this situation, but to simplify notation we do not include $f(\bX; \theta)$ when writing $\K(\cdot)$. The proofs in the Appendix consider this case and make explicit the dependence of $\K(\cdot)$ on $f(\bX; \theta)$, rather than just $\theta$.

\subsection{A Single Split} \label{sec:single_split}
In our experience, there is concern among practitioners that solving (\ref{eq01}) on the complete data is somehow ``cheating," because all of the data are used to build the score while simultaneously estimating risk. The concern is that risk estimators in such a scenario will overstate the usefulness of the score. One way to alleviate this concern is to split the data into two disjoint groups, ${\cal D}_{in}$ and ${\cal D}_{out}$, and use these to estimate $\theta$ and $\beta$ respectively. Let $(\delta_1,\dots,\delta_n)$ be independent and identically distributed Bernoulli$(\pi)$ random variables. In our applications we set $\pi = 1/2$ so that the splits are approximately equal size. We denote the first partition, ${\cal D}_{in}$, by $\delta = 1$, which estimates $\boldtheta$ by solving
\be
0 &=& \sumi \delta_i \Psi(W_i,\bX_i,\bZ_i,\boldtheta). \label{eq03}
\ee
In the second partition, ${\cal D}_{out}$, denoted by $\delta = 0$, we estimate $\boldbeta$ by solving
\be
0 &=& \sumi (1-\delta_i) \K(Y_i,\bX_i,\bZ_i,\boldbeta,\wh{\boldtheta}). \label{eq04}
\ee
Since the two groups are independent, solving (\ref{eq03})--(\ref{eq04}) alleviates concerns about overfitting. This is an analogue to the variable selection procedure of \cite{wasserman2009high}, who also propose a single split, with variable selection conditioned on the data with $\delta = 1$.

\subsection{Many Splits} \label{sec:many_splits}
Section \ref{sec:single_split} give a $\wh{\beta}$ that depend on the particular random sample split. \cite{meinshausen2009} and \cite{dezeure2015high} criticize this and call it a ``p-value lottery" as the p-value can vary from 0 to 1 depending on the split. Instead, in their context, they suggest using multiple data splits to eliminate simulation variability from choosing only a single split. Such an approach would randomly split the data into two parts $b=1,\dots,B$ times to create $B$ partitions of the data.

Define $B$ to be the number of sample splits. For $b=1,\dots,B$ and $i=1,\dots,n$, let $(\delta_{1b},\dots,\delta_{nb})_{b = 1}^{B} $ be independent and identically distributed Bernoulli$(\pi)$ random variables.  If $\delta_{ib} = 1$, then the $i\th$ person is selected into the $b^{th}$ training set \dinb. Then solve
\bse
0 &=& \sumi \delta_{ib}\Psi(W_i,\bX_i,\bZ_i,\boldtheta). \label{eq:psi_b}
\ese
to estimate $\wh{\boldtheta}_b$. The subscript denotes the dependence on the parameter estimate of the $b^{th}$ sample split.

If $\delta_{ib} = 0$, then the $i\th$ person is selected into the test set, \doutb. We then estimate $\wh \boldbeta_b$ by solving
\bse
0 &=& \sumi (1-\delta_{ib}) \K(Y_i,\bX_i,\bZ_i,\boldbeta,\wh{\boldtheta}). \label{eq:kappa_b}
\ese

This process gives $B$ estimates of $\boldtheta$ and $\beta$. We combine them with the sample mean to get $\wh{\boldtheta} = B^{-1}\sumb \wh{\boldtheta}_b$ and $\wh{\boldbeta} = B^{-1}\sumb \wh{\boldbeta}_b$.  Although $\wh \theta_1,\dots,\wh \theta_B$  and $\wh \beta_1,\dots,\wh \beta_B$ are dependent sequences, in the next section we are able to establish the limiting distributions of $\wh \theta$ and $\wh \beta$ as $B \to \infty$ and $n \to \infty$.

\section{Asymptotic Theory} \label{sec:asymptotics}
In this section, we provide asymptotic theory for three cases of sample splitting: $B = 1, 1 < B < \infty$, and $B \to \infty$. We provide asymptotic expansions from which asymptotic normality and the asymptotic covariance matrix can be derived. Hypothesis tests can be performed using Wald test statistics and Normal-based confidence intervals.  However, for testing we recommend the exact tests in \cite{kravitz2019a}.

\subsection{Single Split Asymptotics} \label{sec:single_split_theory}
Suppose there is only a single split, so $B=1$ and in what follows $b = 1$. Define
\bse
\Omega_{nb}(\boldtheta) &=& n^{-1} \sumi \delta_{ib} E\{\partial \Psi(W, \theta) / \partial \theta\trans\};\\
\Lambda_{nb}(\boldbeta,\boldtheta) &=& n^{-1}\sumi (1-\delta_{ib}) E\{\partial {\cal K}(Y, \beta,\theta) / \partial \beta\trans\};\\
\Delta_{nb}(\boldbeta,\boldtheta) &=& n^{-1} \sumi  (1-\delta_{ib}) E\{\partial {\cal K}(Y, \beta,\theta) / \partial \theta\trans\}.
\ese

Using standard M-estimation theory \citep{stefanski2002calculus}, and suppressing the dependence of $\{\Omega_{nb}(\boldtheta),\Lambda_{nb}(\boldbeta,\boldtheta),\Delta_{nb}(\boldbeta,\boldtheta)\}$ on $(\boldbeta,\boldtheta)$, we find that
\be
n^{1/2}(\wh{\theta}_{b} - \theta) &=& -\Omega_{nb}^{-1} n^{-1/2} \sumi  \delta_{ib} \Psi(W_i,\theta) + o_P(1). \label{eq:theta_b}\\
n^{1/2}(\wh{\beta}_{b} - \beta) &=&  -\Lambda_{nb}^{-1} n^{-1/2} \sumi  (1-\delta_{ib}) \K(Y_i, \beta,\theta)\nonumber \\
&& \hskip 10mm +\Lambda_{nb}^{-1} \Delta_{nb} \Omega_{nb}^{-1} n^{-1/2} \sumi  \delta_{ib} \Psi(W_i,\theta) + o_P(1). \label{eq:beta_b}
\ee

Define
\bse
{\cal A}_{ib} = \left[\begin{array}{c}-\Omega_{nb}^{-1} \delta_{ib}\Psi(W_i,\theta)\\
	-\Lambda_{nb}^{-1} (1-\delta_{ib}) \K(Y_i, \beta,\theta)
	+\Lambda_{nb}^{-1} \Delta_{nb} \Omega_{nb}^{-1} \delta_{ib} \Psi(W_i,\theta)
\end{array}\right],
\ese
and define $\wh {\cal A}_{ib}$ as ${\cal A}_{ib}$ with all unknown quantities replaced with their empirical estimates. Then the joint asymptotic covariance matrix of $(\wh{\theta}_{b},\wh{\beta}_{b})$ can be estimated by $n^{-1} S(\wh {\cal A}_{ib})$, where $S(\cdot)$ is the sample covariance.

\subsection{Multiple Split Asymptotics} \label{sec:many_split_theory}
Fix a finite number of splits, $1 < B < \infty$. The final estimates of $\boldtheta$ and $\bbeta$ are the average of the single split estimates from Section \ref{sec:single_split_theory}.  Thus, the expansions of $\theta_b$ and $\beta_b$ in (\ref{eq:theta_b}) and (\ref{eq:beta_b}) can be replaced with their averages over $B$
\bse
n^{1/2}(\wh{\theta}_{b} - \theta) &=& -B^{-1}\sumb  \Big\{ n^{-1/2} \Omega_{nb}^{-1}  \sumi  \delta_{ib} \Psi(W_i,\theta) \Big\} + o_P(1), \label{eq:avg_theta_b}\\
n^{1/2}(\wh{\beta}_{b} - \beta) &=&  -B^{-1} \sumb \Big\{ \Lambda_{nb}^{-1} n^{-1/2} \sumi  (1-\delta_{ib}) \K(Y_i, \beta,\theta)\nonumber \\
&& \hskip 10mm - \Lambda_{nb}^{-1} \Delta_{nb} \Omega_{nb}^{-1} n^{-1/2} \sumi  \delta_{ib} \Psi(W_i,\theta) \Big\} + o_P(1). \label{eq:avg_beta_b}
\ese

Define ${\cal A}_{i} = B^{-1}\sumb {\cal A}_{ib}$, where ${\cal A}_{ib}$ is the same as in Section \ref{sec:single_split_theory}.  Define $\wh{\cal A}_i = \sumb \wh{\cal A}_{ib}$. The joint asymptotic covariance matrix of $(\wh{\theta},\wh{\beta})$ can be estimated consistently by $n^{-1}  S(\wh{{\cal A}}_{i})$.

\subsection{Increasing Number of Splits Asymptotics} \label{sec:infinite_split_theory}
We  assume that $B$ increases at a slower rate than $n$, $B = B_n = o_P(n^{1/2 - a})$ for any $a>0$. Since the user chooses $B$, this is not a restrictive assumption. Since the $\delta_{ib}$ are independent of the rest of the data, $B^{-1} \sumb \delta_{ib} \rightarrow \pi$ and $B^{-1} \sumb (1 - \delta_{ib}) \rightarrow (1 - \pi)$. The asymptotic expansions for $\wh \theta$ and $\wh \beta$ simplify to
\be
n^{1/2}(\wh{\theta} - \theta) &=& -\Omega^{-1} n^{-1/2} \sumi  \pi \Psi(W_i,\theta) + o_P(1); \label{eq:infinite_theta_b} \\
n^{1/2}(\wh{\beta} - \beta) &=& -\Lambda^{-1} n^{-1/2} \sumi  (1-\pi) {\cal K}(Y_i,\beta,\wh{\theta}_b) +  \nonumber \\
&& \hskip 15mm \Lambda^{-1}n^{-1/2} \Delta \Omega^{-1} \sumi \pi \Psi(W_i,\theta) + o_P(1).  \label{eq:infinite_beta_b}
\ee
The asymptotic expansions in (\ref{eq:infinite_theta_b}) and (\ref{eq:infinite_beta_b}) are also the expansion of $\wh \beta$ and $\wh \theta$ using stacked estimating equations with all the entire sample. This is justified in Appendix \ref{sec:asymptotic_equivalence}. This shows that when $\pi=1/2$, the estimates from sample splitting become asymptotically equivalent to estimates from the entire data set and not using sample splitting.

\section{Simulations} \label{sec:simulations}

We simulate from two configurations. We consider a single split ($B = 1$) and $B = 25$ sample splits and set $\pi = 1/2$, so each \dinb\ and \doutb\ are approximately the same size. We express both configurations as regression models to simplify interpretation but they can be expressed as estimating equations.

Both simulations are related to the problem of estimating the ratio of two normal random variables \citep{fieller1932distribution,fieller1954some}; see \cite{wang2015direct} for a recent review and some alternative methods. It is known that there is no confidence interval with guaranteed coverage that has finite length with probability one. It is also known that asymptotic standard errors based on estimating equations and the delta method generally have somewhat less than nominal coverage probabilities except for larger sample sizes. We will see the Fieller phenomenon (under coverage) occur in our simulations.

The first configuration is a set of two linear regression models,
\be
\label{eq:linear_simulation}
W_i = \bX_i \trans \theta + \epsilon_i; \hskip 15mm	Y_i = \beta _0 (\bX_i \trans \theta ) + \epsilon_i.	
\ee
We set $\theta = c(1,1,1) / \sqrt{3}$, $\beta_0 = 1 / \sqrt{3}$,  $\epsilon_i \sim N(0, 0.5)$, and vary $n = 25, 50, 100, 250, 500$. Results are in Table \ref{tab:simulation_linear}. Confidence intervals have approximately 95\% coverage for $n > 250$.

The second configuration is two binary regression models, given as 
\be  \label{eq:logistc_simulation}
\pr(W_i = 1 | \bX_i)  = H( \bX_i \trans \theta);
\hskip 15mm \pr(Y_i = 1 | \bX_i ) &=& H\{ \beta _0 f(\bX_i; \theta ) \} 	,
\ee
where $H(\cdot)$ is the logistic distribution function. We set $f(\bX_i, \theta) = 2 + 1.5 (\bX_i \trans \theta)^2$. The specific $f(\bX, \theta)$ is set to mimic the case study introduced in Section \ref{sec:motivation} and analyzed in Section \ref{sec:data_analysis}. This transformation also rescales the fitted values, although not to the same 0 to 100 range, and contains a nonlinear transformation of the parameters and data. The new scale is arbitrary; we choose the values $2$ and $1.5$, so there are a reasonable number of both 0's and 1's in $ Y$. Again we set $\theta = (1,1,1)\trans / \sqrt{3}$ and $\beta_0 = 1 / \sqrt{3}$ and vary $n = 50, 100, 250, 500$. Results for $B = 1$ and $B = 25$ are in Table \ref{tab:simulation_logistic}, and are much the same as in Table \ref{tab:simulation_linear}, with the Fieller phenomenon being observed. Logistic regression naturally requires larger samples than linear regression to achieve nominal coverage.

\section{Data Analysis} \label{sec:data_analysis}
 
Participants in the NIH AARP Study of Diet and Health were ask to complete a questionnaire to measure physical behaviors, medical history, and risk factors for disease. Around a fifth of the total participants (N = 163,106) responded and met criteria for inclusion. Survey responses were translated to time or energy spent in five aerobic activities, two types of sitting activities, and sleep.
 
 We model each of the physical activity components to be consistent with the existing kinesiology literature. The dose-response relationship between aerobic activity and survival has been established as somewhat concave and non-decreasing \citep{arem2015leisure}, with benefits for overall health which level off with increasing activity.  Sedentary time is known to have a negative effect on overall health \citep{grontved2011television, prince2014comparison}.  Sleep is known to be beneficial in reasonable doses but too much or too little sleep is indicative of poor health	\citep{yin2017relationship}, suggesting a concave, parabolic relationship with overall health. Table \ref{constraints} lists the expected relationships and the marginal models we use to model these relationships. One might consider constraining the parameters to satisfy these relationships, but this was not necessary as unconstrained maximum likelihood estimates already satisfy them.
 
\subsection{Step 1: Developing the Score}
On each $D_{in}$ we fit the model
\be \label{eq:binary_results}
\hbox{$\pr(W_i = 1 | \bX_i, \bZ_i$) } &=& H \Bigg[ \hbox{$\sum_{j = 1}^{5}$}   \Big\{ d_j - \frac{d_j} {1 + (X_{{\rm aerob}_j} /c_j) ^{b_j}} \Big\} + \theta_{{\rm TV}} X_{{\rm TV}} \nonumber \\
	&+& \theta_{{\rm Sit}} X_{{\rm Sit}} + \theta_{{\rm Sleep}, 1}  X_{{\rm Sleep}} + \theta_{{\rm Sleep}, 2} X^2_{{\rm Sleep}} + \bZ_i \trans \theta\Bigg],
	\ee
where $W_i = 1$ indicates survival until the end of the study, $\bX_i$ is a vector containing the 8 physical activity levels of the $i^{th}$ person in the order listed in Table~\ref{constraints} , $\bZ_i$ is a vector of covariates including sex, race, education status, and an intercept, and $H(\cdot)$ is the logistic distribution function. 

The first row of Figure \ref{fig:combined_plots}, shows the fitted marginal models for three types of activity: moderate physical activity, sleep, and television sitting. The curves match their intended functional form: moderate physical activity is concave and increasing, sleep is concave, and television sitting is decreasing.

\subsection{Step 2: Rescaling the Score}

The fitted values from (\ref{eq:binary_results}) are the logits of the effects of the physical activities on survival. We want to rescale the logits from the physical activity covariates so that their sum is between 0 and 100 with 0 being highest risk and 100 being lowest risk.  While we fit model (\ref{eq:binary_results}) with the additional covariates $\bZ$ to prevent confounding due to demographic information, we do not include the fitted values $\bZ_i \trans \; \wh{\theta}$ in the score. 
	
\begin{figure}[ht]
\begin{center}
\includegraphics[width = .9 \textwidth]{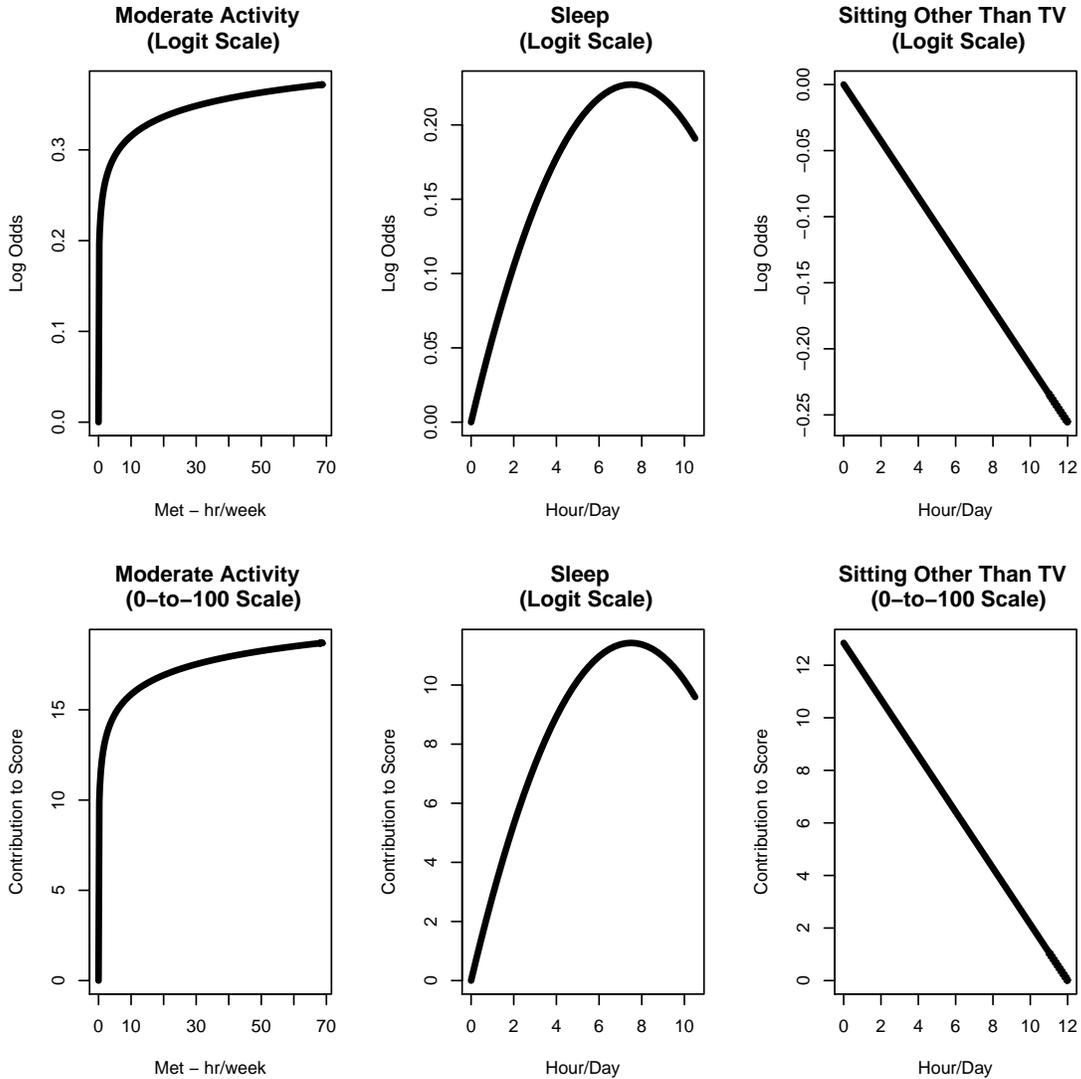}
\caption{\baselineskip=12pt Three of the 8 marginal model plots from binary regression model. The first row shows the marginal models in the original scale and the second rows shows them in a 0-to-100 scale. Moderate activity is modeled to be concave and increase, sleep is modeled to be concave, and television sitting is modeled to be decreasing.  To put the marginal models on a 0-to-100 scale, first make them positive by adding the absolute value of the minimum of each function. Then we rescale the functions so the maximum value of each of the function jointly sum to 100.}
\label{fig:combined_plots}
\end{center}
\end{figure}

To rescale the logits, we first force all the physical activity marginal models to be positive by adding the absolute value of the minimum fitted value. For example, in the top row of Figure \ref{fig:combined_plots} we see that the marginal model for non-TV sitting is negative for any amount of non-TV sitting greater than 0. The function has a minimum of $-0.25$ at around 12 hours per day of non-TV sitting.  By adding $-0.25$ to the fitted values, we can force this function to always be positive. 
	
Next, we sum the maximum value obtained by each of the, now positive, marginal models and denote this with $T$. We  transform the fitted values with
\be \label{eq:score}
\frac{T}{100} \Bigg[ \hbox{$\sum_{j = 1}^{5}$}   \Big\{ d_j - \frac{d_j} {1 + (X_{{\rm aerob}_j} /c_j) ^{b_j}} \Big\} + \theta_{{\rm TV}} X_{{\rm TV}} \nonumber \\
+ \theta_{{\rm Sit}} X_{{\rm Sit}} + \theta_{{\rm Sleep}, 1}  X_{{\rm Sleep}} + \theta_{{\rm Sleep}, 2} X^2_{{\rm Sleep}} \Bigg]
\ee
which puts the fitted values from physical activity on a scale from 0 to 100.  We refer to the rescaled marginal models as the contributions to the total score. Three examples of these rescaled marginal models are shown in the bottom row of Figure \ref{fig:combined_plots}. The contribution to the total score is given on the y-axis. Moderate activity, for example, accounts for up to 15 points of the total score of 100. Table \ref{tab:total_score} has an example of the physical activity score using the results from (\ref{eq:score}) on one particular split of the data. Table \ref{tab:total_score} lists the 8 physical behaviors with their relative contribution to a score of 100 and the criteria for receiving a perfect score in each criteria.

\subsection{Step 3: Risk Prediction Based on the Score}

We denote the physical activity score created in the previous section with $f(\bX, \theta)$.  We then estimate the relationship between the physical behavior score and mortality using a Logistic Regression model on \doutb 
\be \label{eq:results_logistic}
\pr(Y_i = 1 | \bX_i, \bZ_i ) &=& H\{ \beta _0 f(\bX_i; \theta ) + \bZ_i \trans \beta \},
\ee
where $Y_i = 1$ denotes mortality at any point during the study.

\begin{figure}[ht]
\begin{center}
\includegraphics[width = .9 \textwidth]{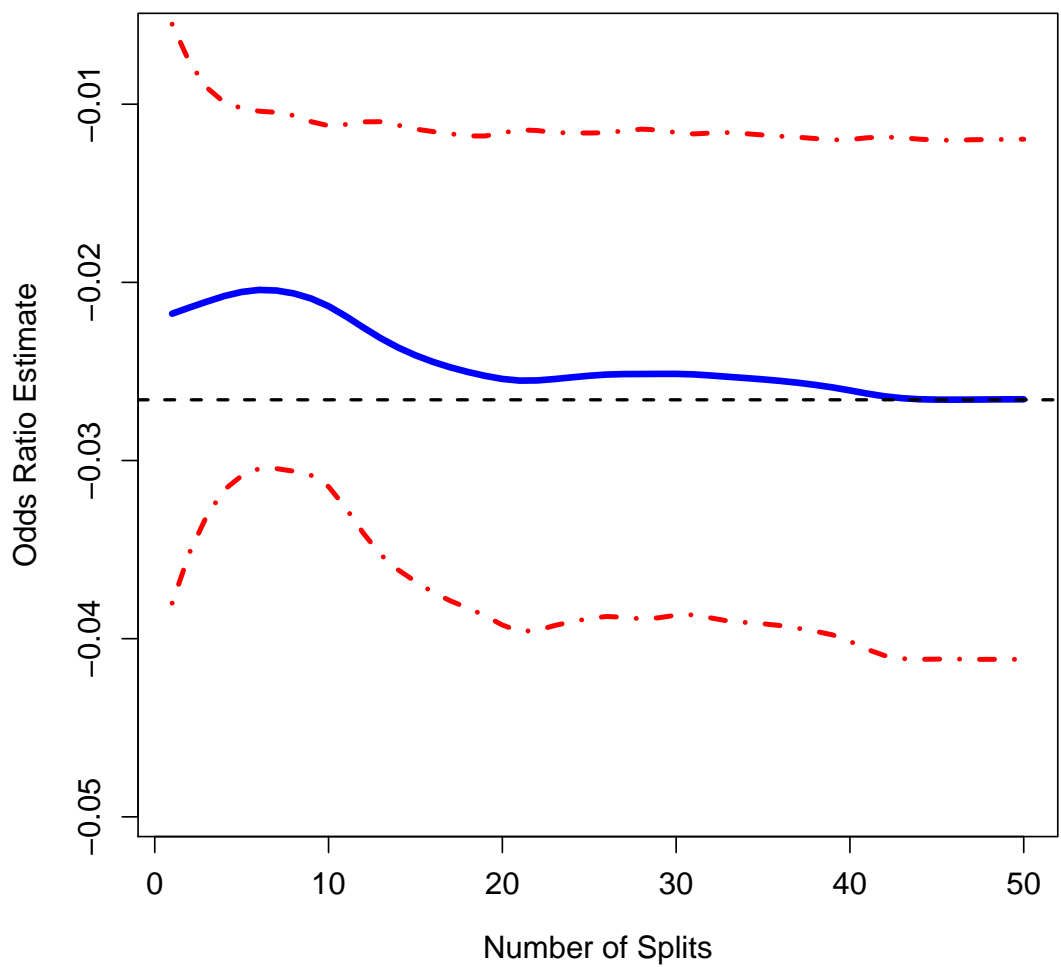}
\caption{\baselineskip=12pt The solid blue line denotes the estimate of $\beta$,
the odds ratio of the physical activity score, as the number of sample splits increases. A 95\% confidence interval is shown with the red dash-dotted lines. The final estimate of $\beta$, averaged over all 50 splits, is shown with a dashed black line at $-0.026$. The confidence intervals and parameter estimates are plotted using a smoothing spline.}
\label{fig:hazard_ratio}
\end{center}
\end{figure}

A plot of the estimates can be seen in Figure \ref{fig:hazard_ratio}. The solid black line shows the limiting value of $\wh \beta_0$ with an increasing number of sample splits. For a particular number of sample splits, $B$, the solid blue curve shows $B^{-1} \hbox{$\sum_{b=1}^{B}$}\wh \beta^{(b)}$. The dashed red curves are 95\% confidence intervals. The overall mean of all these estimates is $\wh \beta = -0.026$ and is calculated with $\wh \beta = 50^{-1} \hbox{$\sum_{b=1}^{50}$} \wh \beta^{(b)}$. It is denoted with a dashed black  line.

There is variability when using a small number of number of splits. This is visible in Figure \ref{fig:hazard_ratio} for small values of $B$. Despite this, $\wh \beta_0$ changes little. The difference between the estimates of $\beta_0$ when $B = 1$ and $B = 50$ is only 0.004 and the confidence intervals are almost entirely overlapping.

\section{Discussion} \label{discuss}
We explored sample splitting to valid a fitted model when a second dataset is unavailable and validation is done with the same data used for model fitting.  This scenario is of interest to practitioners who want a formal distinction between the data used for model building and  data used for validation (e.g., hypothesis testing). 

In practice, there are two ways to evaluate a marginal score. One may treat the newly created score as a fixed quantity used as a predictor in a regression model. This ignores the variability due to estimating this score and underestimates uncertainty. Instead, we consider the score as a function of unknown parameters which are estimated with error. This results in consistent parameter estimate which have higher variability than had the score been considered fixed.

We provided asymptotic expansions for a fixed sample split and a mean-aggregated multiple split which can be used for hypothesis testing and asymptotic confidence intervals. In particular we are able to provide confidence intervals for a single split of the data.

As mention in Section \ref{sec:introduction},  \cite{kravitz2019a} used sample splitting to obtain an exact test of $\beta_0 =0$. Since a primary question is whether the index has predictive power, this is an important result.  It is likely that researchers will also want a confidence interval for $\beta_0$, since the size of the effect is of considerable interest.  To the best of our knowledge, no exact confidence interval is available.  We developed a confidence interval based on the theory of estimating equations, which is asymptotically correct and close to nominal coverage probability when $n \ge 250$ (Table \ref{tab:simulation_logistic}).
In our application, $n$ is much larger than this.

We find that as the sample size ($n$) and number of splits ($B$) approach infinity, the sample splitting estimator converges in probability to the stacked estimating equations estimator.
This interesting result implies that confidence intervals do not require sample splitting, but instead could be applied when the entire dataset is used both to develop the index and to test for an effect of the index.  However, we recommend that sample splitting and the exact test in \cite{kravitz2019a} be used to test whether $\beta_0$ is zero.

\baselineskip=16pt
\section*{\Supp}

The R programs used in the simulations of Section \ref{sec:simulations} and the data analysis of Section \ref{sec:data_analysis} are available by request or on Github at https://github.com/kravitzel/Sample\_Splitting

We do not have permission to distribute the actual data involved in Section \ref{sec:data_analysis}: such data can be obtained via a data transfer agreement with the National Cancer Institute, see https://epi.grants.cancer.gov/Consortia/cohort\_projects.html.

\section*{Acknowledgments}
Carroll's research was supported by a grant from the National Cancer Institute (U01-CA057030).

\baselineskip=14pt
\baselineskip=14pt
\bibliographystyle{biomAbhra}
\bibliography{asymptotics_July2019}

\newcommand{\Appendix}{\appendix\def\thesection{Appendix~\Alph{section}}\def\thesubsection{\Alph{section}.\arabic{subsection}}}
\section*{Appendix: Sketch of Technical Arguments}
\begin{appendix}
\Appendix
\renewcommand{\theequation}{A.\arabic{equation}}
\renewcommand{\thesubsection}{A.\arabic{subsection}}
\setcounter{equation}{0}

\subsection{Assumptions} \label{sec:assumptions}

Suppose $\wh{\theta}$ is the solution to
\be \label{eq:EE_theta}
0 = n^{-1} \sumi \Psi(W_i, \bX_i,\bZ_i,\theta),
\ee	
where $\Psi(\cdot)$ is a known influence function that does not depend on $i$ and is continuous and twice differentiable with respect to $\theta$ and $\beta$. 
Suppose  $\wh{\beta}$ is the solution to 	
\bse
0 = n^{-1} \sumi \K\{ Y_i, \beta,  f(\bX,\wh{\theta}) \},
\ese
where $\K(\cdot)$, as in (\ref{eq:EE_theta}), is a known function which does not depend on $i$, $\K(\cdot)$ twice differentiable with respect to $\beta$, and differentiable with respect to $f(\bX, \theta)$. Assume $f(\bX, \theta)$ is  differentiable with respect to $\theta$.

We assume the estimators from different splits are exchangeable and unbiased. That is, $E(\wh \theta_k) = E(\wh \theta_{\ell}) = \theta$ and $E(\wh \beta_k) = E(\wh \beta_{\ell}) = \beta$ for all $k, \ell$.

\subsection{Taylor Expansions from Sections \ref{sec:single_split_theory} and \ref{sec:many_split_theory}} \label{sec:taylor_expansions}

For notational simplicity, in what follows, the covariate vectors $\bX_i$ and $\bZ_i$ and not included in $ \Psi(W_i, \bX_i,\bZ_i,\theta)$.
Since $\wh \theta_b$ and $\wh \beta_b$ are asymptotically linear they have expansions
 \be \label{eq:split_theta_linearity}
n^{1/2}(\wh{\theta}_{b} - \theta) &=&  -\Omega_{nb}^{-1} n^{-1/2} \sumi  \delta_{ib} \Psi(W_i, \theta)  + o_P(1). \\
\label{eq:split_beta_linearity}
\hskip 15mm
n^{1/2}(\wh{\beta}_b - \beta) &=& -\Lambda^{-1}_{nb} n^{-1/2} \sumi (1-\delta_{ib}) \K_{ib}\{Y_i, f(\bX_i,\wh{\theta}_{b}) \bZ_i ,\beta\}+ o_P(1).
\ee
Taking a Taylor expansion of (\ref{eq:split_beta_linearity}) around $\theta$  we have
\bse
&& n^{-1/2} \sumi (1-\delta_{ib}) \K\{Y_i, f(\bX_i,\wh{\theta}_{b}),\beta\} \nonumber \\
&& \hskip 15mm= n^{-1/2} \sumi (1-\delta_{ib}) \K\{Y_i, f(\bX_i,\theta),\beta\} \nonumber \\
&& \hskip 15mm + n^{-1} \sumi (1-\delta_{ib}) \K_{f}\{Y_i, f(\bX_i,\theta),\beta\}f\trans _{\theta}(\bX_i,\theta) n^{1/2}(\wh{\theta}_{b}-\theta) + o_P(1)  \label{eq:taylor_expand_2}.
\ese
Define
\bse \label{eq:delta_definition}
\Delta_{nb} = n^{-1} \sumi (1-\delta_{ib}) \K_{f}\{Y_i, f(\bX_i,\theta),\beta\}f\trans _{\theta}(\bX_i,\theta) .
\ese
Taking a Taylor expansion from (\ref{eq:taylor_expand_2}) and plugging (\ref{eq:split_theta_linearity}) in place of $n^{1/2}(\wh \theta_b - \theta)$ we get
\bse
 n^{1/2}(\wh{\beta}_b - \beta) &=& -\Lambda^{-1}_{nb} n^{-1/2} \Big[ \sumi \
 (1-\delta_{ib}) \K\{Y_i, f(\bX_i,\theta),\beta\}  \nonumber  \\
\hskip 15mm && - \ \Delta_{nb} \Omega_{nb}^{-1} \sumi \delta_{ib} \Psi(W_i, \theta) \Big] + o_P(1).
\ese
The asymptotic expansion of $\wh \theta = B^{-1} \sumb \theta_b$ and $\wh \beta = B^{-1} \sumb \beta_b$ are the averages over $B$ of the previous expansion,
\be
n^{1/2}(\wh{\theta}_{b} - \theta) &=& -B^{-1} \sumb  \Omega_{nb}^{-1} n^{-1/2} \sumi  \delta_{ib} \Psi(W_i,\theta) + o_P(1), \label{eq:sup_avg_theta_b}\\
\hskip 15mm
n^{1/2}(\wh{\beta}_{b} - \beta) &=&  -B^{-1} \sumb \Lambda_{nb}^{-1} n^{-1/2} \Big\lbrack \sumi  (1-\delta_{ib}) \K\{Y_i, f(\bX, \theta), \beta\}\nonumber \\
&& \hskip 15mm - \Delta_{nb} \Omega_{nb}^{-1} \sumi  \delta_{ib} \Psi(W_i,\theta) \Big\rbrack + o_P(1). \label{eq:sup_avg_beta_b}
\ee

\subsection{Theory for Infinite Splits from Section \ref{sec:infinite_split_theory}}

We note of two technical details. First
\bse
\wh{\Omega}_{nb}(\wh{\boldtheta}_b) = \pi E\{\partial \Psi(W,\theta) / \partial \theta\trans\} + O_P(n^{-1/2}) = \Omega  + O_P(n^{-1/2}) \label{eq:omega_expectation} \\ \hskip 10mm
\wh{\Lambda}_{nb}(\wh{\boldtheta}_b, \wh{\boldbeta}_b) = (1 - \pi) E\{\partial \K(Y,\beta, \theta) / \partial \beta\trans\} + O_P(n^{-1/2}) = \Lambda  + O_P(n^{-1/2}) \label{eq:lambda_expectation}. \\ \hskip 10mm
\wh{\Delta}_{nb}(\wh{\boldtheta}_b, \wh{\boldbeta}_b)  =  (1- \pi) E \lbrack \partial {\cal K}\{Y, \beta, f(\bX; \theta) \} / \partial \theta\trans \rbrack + O_P(n^{-1/2}) = \Delta + O_P(n^{-1/2}).
\ese
Also, as in \cite{carroll1996asymptotics}, the $o_P(1)$ terms in the expansion in (\ref{eq:split_theta_linearity}) and (\ref{eq:split_beta_linearity}) are  $O_P(n^{-1/2+a})$ for any $a>0$.
Combining these, we have,
\bse
n^{1/2}(\wh{\theta}_{b} - \theta) &=& \{-\Omega^{-1} + O_P(n^{-1/2})\} n^{-1/2} \sumi  \delta_{ib} \Psi(W_i,\theta) + O_P(n^{-1/2+a}) \nonumber \\
\hskip 15mm &=& -\Omega^{-1}n^{-1/2} \sumi  \delta_{ib} \Psi(W_i,\theta) + O_P(n^{-1/2+a})  \label{ams02} \\
 n^{1/2}(\wh{\beta}_b - \beta) &=& -\Lambda^{-1} n^{-1/2} \sumi
\lbrack (1-\delta_{ib}) \K\{Y_i, f(\bX_i,\theta),\beta\} \nonumber \\
\hskip 15mm &&+ \Lambda^{-1} \Delta \Omega^{-1} \delta \Psi(Y_i, \theta) \rbrack + O_P(n^{-1/2 + a}).
\ese
Since the $\delta_{ib}$ are independent of the data, $B^{-1} \sumb \delta_{ib} = \pi+o_P(1)$, and since $B = B_n = o_P(n^{1/2 - a})$ by assumption, the expansion in (\ref{eq:sup_avg_theta_b}) and (\ref{eq:sup_avg_beta_b}) become
\be
n^{1/2}(\wh{\theta} - \theta) &=& -\Omega^{-1} n^{-1/2} \sumi  \pi \Psi(W_i,\theta) + o_P(1) \nonumber \\
n^{1/2}(\wh{\beta} - \beta) &=&  -\Lambda^{-1} n^{-1/2} \sumi  (1-\pi) \K(Y_i, \beta,\theta)\nonumber \\
&& \hskip 10mm +\Lambda^{-1} \Delta  \Omega^{-1} n^{-1/2} \sumi  \pi \Psi(W_i,\theta) + o_P(1). \label{eq:b_infinity_beta}
\ee

\subsection{Asymptotic Equivalence to Stacked Estimating Equations} \label{sec:asymptotic_equivalence}

Both $\theta$ and $\beta$ can be estimated by the stacked estimating equations instead of the sample splitting of the previous sections. The stacked estimating equations are given by,
\bse
\sumi\left(
\begin{array}{c}
	\Psi(Y_i,\theta) \\
	\K\{W_i, f(\bX_i,\theta),\beta\}
\end{array}
\right)
=
\left(
\begin{array}{c}
	0 \\
	0
\end{array}
\right).
\ese
Define
\bse
\Phi =
\left(
\begin{array}{cc}
\Phi_{11} & 0 \\
\Phi_{21} & \Phi_{22}
\end{array}
\right)
= E
\left(
\begin{array}{cc}
\partial   \Psi(Y_i,\theta)/ \partial \theta\trans & \partial   \Psi(Y_i,\theta)/ \partial \beta\trans \\
\partial  \K\{W_i, f(\bX_i,\theta),\beta\}/ \partial \theta\trans & \partial \K\{W_i, f(\bX_i,\theta),\beta\}/ \partial \beta\trans
\end{array}
\right)
\ese
and
\bse
\Phi^{-1} = \left(
\begin{array}{cc}
 \Phi_{11}^{-1} & 0\\
 - \Phi_{22}^{-1}\Phi_{21}\Phi_{11}^{-1} & \Phi_{22}^{-1}
\end{array}
\right).
\ese

Then it follows that the stacked estimating equations estimator of $\beta$ has influence function
\be \label{eq:stacked_EE_linear}
{\cal G}_i  = \Big[ \Phi_{22}^{-1} \K\{W_i, f(\bX_i,\theta),\beta\}  - \Phi_{22}^{-1} \Phi_{21} \Phi_{11}^{-1} \Psi(Y_i,\theta)\Big]  + o_P(1) \label{eq:stacked_influence}
\ee
Since $\Phi_{22} = (1 - \pi)^{-1} \Lambda$, $\Phi_{11} = \pi^{-1} \Omega$ and $\Phi_{21} = (1 - \pi)^{-1} \Delta$, then (\ref{eq:stacked_EE_linear}) simplifies to (\ref{eq:b_infinity_beta}). Therefore as $B, n \rightarrow \infty$ and when $\pi = 1/2$, the sample splitting estimator becomes equivalent to the estimator from stacked estimating equations.

\end{appendix}

\baselineskip=12pt
\clearpage\pagebreak\newpage
\thispagestyle{empty}

\begin{table}[ht]
	\begin{center}
		\begin{tabular}{|l|c|c| }
			\hline\hline
			\textbf{Activity} & \textbf{Expected Relationship} & \textbf{Marginal Model} \\ \hline
			Vigorous Activity & Concave Increasing & 3-parameter Logistic \\*[-.60em]
			Moderate Activity& Concave Increasing & 3-parameter Logistic  \\*[-.60em]
			Light Household Activity & Concave Increasing & 3-parameter Logistic  \\*[-.60em]
			MVPA Household Activity & Concave Increasing& 3-parameter Logistic   \\*[-.60em]
			Weight Training & Concave Increasing & 3-parameter Logistic  \\*[-.60em]
			Hours Sitting Other than TV &  Decreasing & Linear \\*[-.60em]
			Hours of TV Sitting & Decreasing  & Linear\\*[-.60em]
			Hours of Sleep & Concave  & Quadratic \\
			\hline\hline
		\end{tabular}
		\caption{\baselineskip=12pt Each of the 8 physical activity variables with their expected relationship and marginal model when predicting survival. The expected relationship is derived from the physical activity.}
\label{constraints}
\end{center}
\end{table}

\begin{table}[ht]
\begin{center}
\begin{tabular}{|l|cc|}
\hline\hline
 & \multicolumn{2}{c|}{Linear} \\*[-.60em]
        & $B = 1$ & $B = 25$ \\ \hline
$n = 50$  & 92.60 & 90.45 \\*[-.60em]
$n = 100$ & 93.70 & 92.85\\*[-.60em]
$n = 250$ & 94.90 & 94.15 \\*[-.60em]
$n = 500$ & 94.70 & 95.15 \\*[-.60em]
$n = 1000$ & 95.10 & 94.95 \\
\hline\hline
\end{tabular}
\caption{\baselineskip=12pt Results of the simulation in Section \ref{sec:simulations} using the linear model (\ref{eq:linear_simulation}). Coverage of asymptotic 95\% confidence intervals is provided for $B = 1$ and $B = 25$ sample splits.}
\label{tab:simulation_linear}
\end{center}
\end{table}

\begin{table}[ht]
\begin{center}
\begin{tabular}{|l|cc|}
\hline\hline
 & \multicolumn{2}{c|}{Logistic} \\*[-.60em]
        &  $B = 1$ & $B = 25$ \\ \hline
$n = 50$  & $\cdot$ & $\cdot$ \\*[-.60em]
$n = 100$  & 88.95 & 85.40\\*[-.60em]
$n = 250$  & 93.85 & 90.40\\*[-.60em]
$n = 500$  & 93.40 & 94.00\\*[-.60em]
$n = 1000$  & 94.20 & 94.55 \\
\hline\hline
\end{tabular}
\caption{\baselineskip=12pt Results of the simulation in Section \ref{sec:simulations} using the logistc model (\ref{eq:logistc_simulation}). Coverage of asymptotic 95\% confidence intervals is provided for $B = 1$ and $B = 25$ sample splits. Results from the logistic model are unstable at $n=50$ and are omitted.}
\label{tab:simulation_logistic}
\end{center}
\end{table}

\begin{table}[ht]
\begin{center}
    \begin{tabular}{|ccc|}
    \hline\hline
    Component                & Contribution to Total & Criteria for Maximum                           \\ \hline  
    Vigorous Activity        & 10                    & \textgreater{}20 MET-hrs/wk                     \\*[-.60em]
    Moderate Activity        & 30                    & \textgreater{}50 MET-hrs/wk \\*[-.60em]
    Light Household Activity & 3                     & \textgreater{}3 MET-hrs/wk                       \\*[-.60em]
    MVPA Household Activity  & 25                    & \textgreater{}20 MET-hrs/wk                      \\*[-.60em]
    Weight Training          & 2                     & \textgreater{} 2 MET-hrs/wk                       \\*[-.60em]
    Sitting Other than TV    & 6                     & \textless{} 3.5 hours                             \\*[-.60em]
    Hours of TV Sitting      & 14                    & \textless{} 2  hours                              \\*[-.60em]
    Hours of Sleep           & 10                     & 7.5 hours \\ \hline                                  
    Total                    & 100                   & \\  \hline\hline                       
    \end{tabular}
    	\caption{\baselineskip=12pt Example physical activity score developed using half of the data. We fit the binary regression model from Section \ref{sec:motivation} and rescale the fitted values of the physical behaviors to be between 0 and 100. The middle column gives the proportion of the total score of 100 that each component can contribute. The third column gives the criteria for receiving the maximum score for each component. MET = metabolic equivalent.}
\label{tab:total_score}
    \end{center}
    \end{table}	

\begin{table}[ht]
\begin{center}
\begin{tabular}{|c|cccc|}
\hline\hline
  &$B = 1$   & $B = 10$ & $B = 25$  & $B = 50$   \\ \hline
  
$\wh \beta$ & $-0.022$   & $-0.020$   & $-0.025$  & $-0.026$  \\*[-.60em]
95\% CI & $(-0.040, -0.0049)$  & $(-0.030, -0.011)$ & $(-0.038, -0.011)$ & $(-0.041, -0.012)$ \\
\hline\hline
\end{tabular}
\caption{\baselineskip=12pt Results of the data analysis in Section \ref{sec:data_analysis}. Estimated coefficients and 95\%  confidence intervals (denoted with CI) from varying numbers of sample splits.}
\label{tab:real_data_results}
\end{center}
\end{table}

\end{document}